# MALICIOUS WEB SCRIPT-BASED CYBER ATTACK PROTECTION TECHNOLOGY

JongHun Jung, Hwan-Kuk Kim, Soojin Yoon

Internet Incidents Response Architecture Team
Korea Internet & Security Agency
{`jjh2640, rinyfeel, sjyoon`}@kisa.or.kr

## ABSTRACT

*Recent web-based cyber attacks are evolving into a new form of attacks such as private information theft and DDoS attack exploiting JavaScript within a web page. These attacks can be made just by accessing a web site without distribution of malicious codes and infection. Script-based cyber attacks are hard to detect with traditional security equipments such as Firewall and IPS because they inject malicious scripts in a response message for a normal web request. Furthermore, they are hard to trace because attacks such as DDoS can be made just by visiting a web page. Due to these reasons, it is expected that they could result in direct damages and great ripple effects. To cope with these issues, in this article, a proposal is made for techniques that are used to detect malicious scripts through real-time web content analysis and to automatically generate detection signatures for malicious JavaScript.*

## 1. INTRODUCTION

Recent introduction of Ajax and HTML5 technologies has enabled dynamic representation of web content, providing compatibility between a client and a server in web environment. However, the efforts to deal with new security vulnerabilities in these technologies, such as the awareness, countermeasure technology development, and standardization are still insufficient. In particular, web-based attacks using malicious scripts can bypass traditional security equipments, such as IDS, IPS and Web Firewall, because, unlike conventional malicious code attacks, they do not download an executable file directly, but they still can be made by combining normal built-in APIs in JavaScript. And also, it is getting harder to detect these attacks as they employ traffic encryption and script obfuscation. The Figure 1 illustrates how a DDoS attack can be made with JavaScript just by accessing a web page. Furthermore, Due to the October 2014 official standard of the W3C html5 is confirmed, the introduction of html5 is becoming accelerate.it is expected that cyber attacks exploiting vulnerabilities of new tags and APIs will grow rapidly.

In this article, a proposal is made for techniques that are used to detect malicious scripts through collection of HTTP Web traffics and static/dynamic analysis, and to generate a detection signature automatically. Chapter 2 shows trends in related studies. Chapter 3 describes techniques that are used to collect and analyze web content for detection of malicious JavaScript. Chapter 4 describes more compact techniques that are used to generate a detection signature automatically with less false positive rate. Finally, Chapter 4 concludes the article.

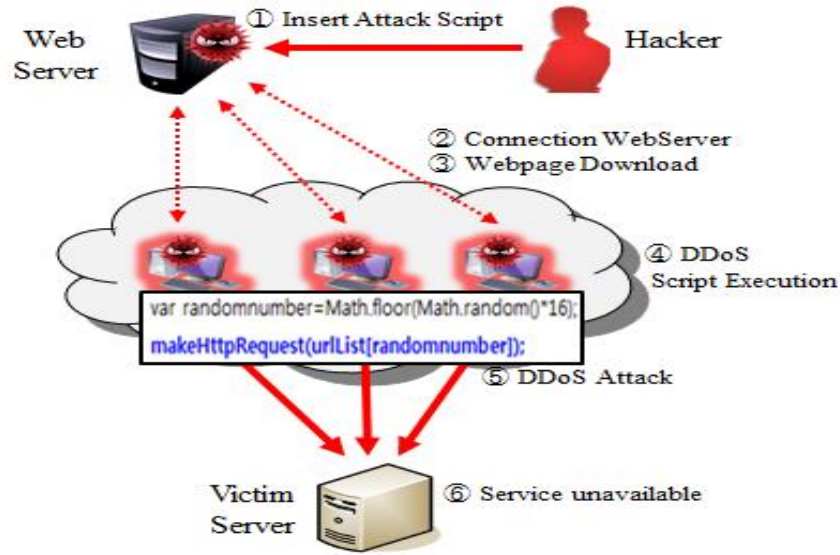

Figure 1. A DDoS attack using JavaScript code

## 2. RELATED WORK

### 2.1. WebShiled

Using a modified browser that consists of DOM API, HTML/CSS Parser and JavaScript Engine only, parse web content in proxy, turn it into the form of DOM Structure, and store it. Send the DOM Structure in the string format to the client. Send a script to the client only if it turns out that the script is safe after running it in the modified browser. However, prevention of exploitation of new vulnerabilities in HTML5 is insufficient.

### 2.2. A signature for Malware Detection

The method of Automatic generation of a signature for malware or worm can be divided into 5 categories: vulnerability-based, content-based, content-shifting, semantic-aware and honeypot-based. Among these, the content-based is the one that is proposed in this article.

In the content-based method, a signature target set is determined based on traffic and the same malicious behavior, and then a signature is generated based on the content.

A content-based signature [1] can be divided into Longest Common Substring, Longest Common Subsequence, Conjunction Signature and Bayes Signature. For the Longest Common Substring and Longest Common Subsequence, one retrieves the longest common substring and longest common subsequence respectively from the target set. For the Conjunction Signature, one uses a set of strings that appear in all targets, as a signature. For Bayes Signature, one checks whether a string in a sample appears in the malicious, and then determines whether the sample is malicious or not based on the percentage of malicious strings.

### 2.3 BIRCH (Balanced Iterative Reducing and Clustering using Hierarchies)

BIRCH is an algorithm for hierarchical clustering for a large database. BIRCH allows addition of a new value in a clustered tree as a new entity is added, eliminating the need of re-clustering.

BIRCH creates a CF (Clustering Feature) tree that has distance information for all leaves under a single node. As a new entity is added, it searches for the closest node. It adds the entity to the cluster of the closest node if the distance is same or shorter than threshold, or creates a new cluster and adds the entity to it if the distance is same or longer than threshold.

## 3. WEB CONTENT ANALYSIS TECHNOLOGY

For real-time detection of malicious JavaScript, one collects HTTP traffics by configuring a proxy server, and parses a HTML document and crawls a link to external resource in order to generate content for analysis. One performs static analysis, such as pattern-matching of web content, and dynamic analysis, such as checking whether obfuscated or not and the HTML5 tag percentage, to determine if the content is malicious. If a malicious script is found, remove the function that actually causes malicious behaviors before sending the script to the client. The Figure 2 shows the proposed system architecture that can be used to detect malicious scripts at network level.

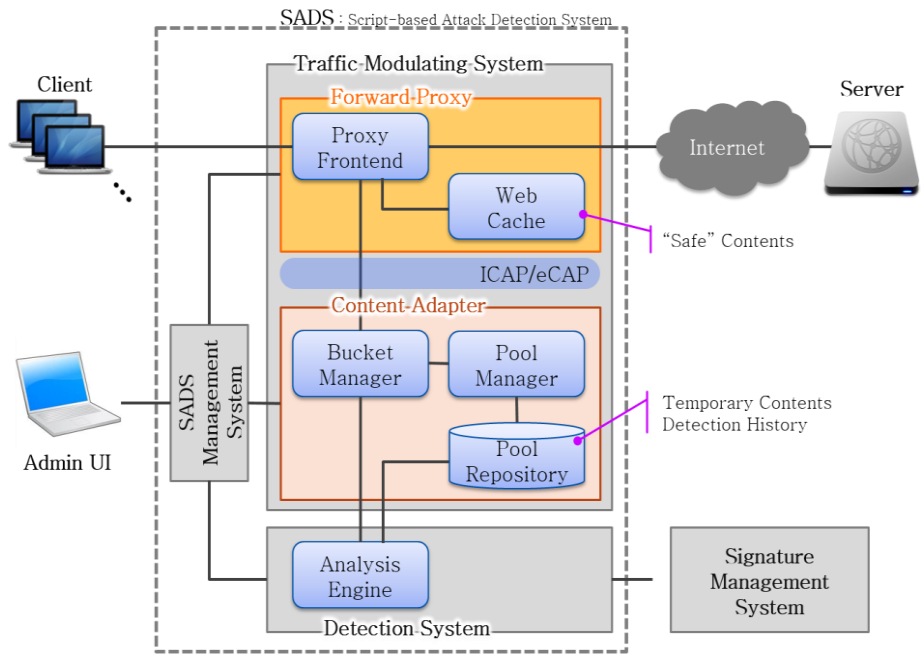

Figure2. System Architecture for Real-time Detection of Malicious Scripts

It consists of modules: i) Forward-Proxy, ii) Web Content Generation Module, iii) Analytics Engine (Static/Dynamic Analysis), and iv) System Control Module.

### 3.1 Forward-Proxy and Content Adapter

For collection of web traffics, Squid-Proxy Server is configured in the in-line form between clients and Web Server, where all HTTP Request and Response packets are collected and Internet Content Adaptation Protocol (ICAP) is used to pass the received HTTP traffics to Web Content Control Module. Then, Web Content Control Module extracts the external JavaScript link data contained in the document, using HTML Parser received from Proxy Server, and collects resources for the link with a separate crawler to generate web content for analysis.

## 3.2 Web Content Analysis

The term 'web content' refers to the entire document that includes both a HTML document and external resources. As shown in the Figure 3, web content goes through the fast static analysis process that performs pattern-matching based on Yara-RuleSet[2]. However, because some sources, such as those obfuscated, require additional analysis, they go through the dynamic analysis process that uses Rihno Browser Engine to run the script and extract call trace data for detection.

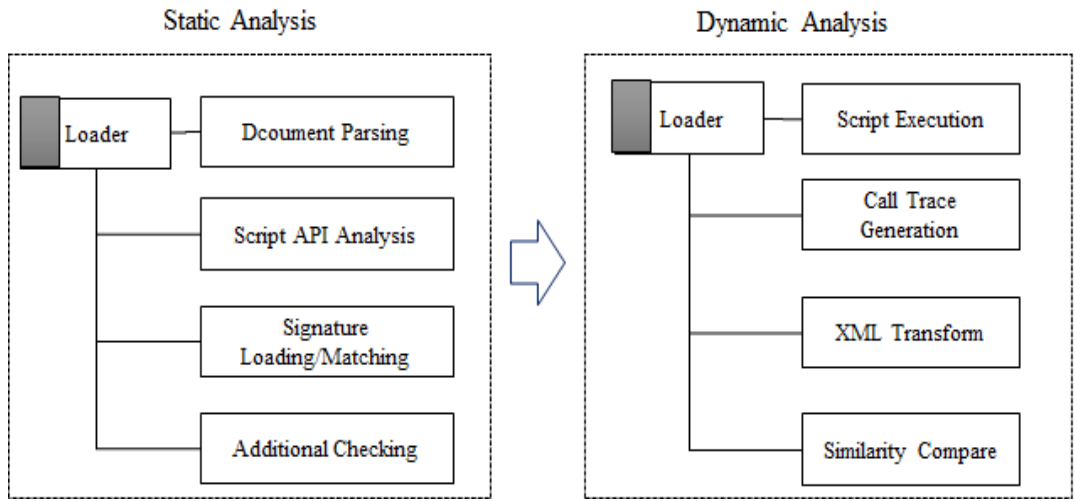

Figure3. Configuration of Web Content Analytics Engine

An input data set is in the format of JSON that consists of a HTML document, external JavaScript, and meta data (IP, port, protocol, domain, etc.). First, extract the primary key token to classify the type of the malicious behavior. Table 1 shows summary of basic keywords contained in each malicious behaviour

Table 1. Examples of Basic Keywords for Each Malicious Behavior

| Malicious Type | | The Keyword |
|---|---|---|
| DoS Attack | HashDoS | setInterval, open, send, ActiveXObject, XMLHTTP, XMLHttpRequest |
| | XML HttpObject DoS | |
| Scan Attack | Network Scan | open, ActiveXObject, XMLHTTP, XMLHttpRequest, Date, readyState |
| | Port Scan | |
| Geolocation | | coords, getCurrentPosition |
| Web Socket | | parse, eval, WebSocket, JSON, send |
| Web Worker DDoS | | postMessage, Worker, XMLHttpRequest, open, send |

Look up the signature for a malicious behavior and then perform signature-matching check to determine whether it is malicious or not.

Additionally, score the JavaScript obfuscation and the percentage of HTML5 new tag usage in the entire document, and then perform dynamic analysis if the score is the same or above the predetermined level. JavaScript obfuscation check is performed because most of malicious JavaScript codes are obfuscated, and it is hard to determine whether it is malicious just by doing signature–matching during static analysis. The Figure 4 illustrates process of the JavaScript obfuscation check[3]. As these are main characteristics of the obfuscated JavaScript, if a special character in a JavaScript string is frequently used, if there is a string with abnormal length, or if the entropy score of characters in the JavaScript is low, score them and if the total score is the same or above the cutoff, consider it obfuscated and perform dynamic analysis additionally.

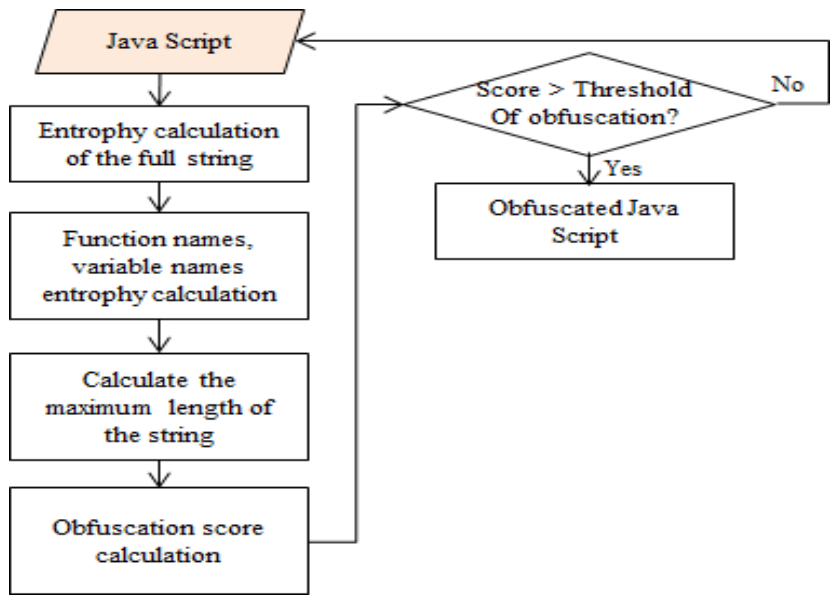

Figure4. JavaScript Obfuscation Analysis Process

The table 2 illustrates the detection result of obfuscated JavaScript. Ngram (frequently used special characters), Word Size (string maximum length), the entropy by adjusting the weights of (JavaScript function name, variable name, Entropy analysis) were detected for the obfuscated script. When it detects a real-obfuscated script was confirmed that at least 87% detection rate.

Table 2. Obfuscated JavaScript detection results

| Obfuscated source | Samples | Weight (Ngram:1.0, Entropy:1.0, Wordsize:1.0) |
|---|---|---|
| Base62 (Norma) packer encode method | 45 | 38(Detection) |
| A packed encode method | 45 | 38(Detection) |
| As Base64 encode method | 10 | 10(Detection) |

| Etc | 24 | 22(Detection) |
|---|---|---|
| sum | 124 | 108(87.1%) |

And also, the usage of HTML5 tags is checked to detect malicious scripts such as jacking or cross-site scripts exploiting new tags of HTML5 (Canvas, Audio, Video). It has been arranged to perform dynamic analysis if a weight for each HTML5 tag is applied and the score is the same or above the predetermined level.

During dynamic analysis in real world situation, a malicious JavaScript is executed using open source-based Rhino JavaScript Engine with a built-in sandbox, and JavaScript API Call Trace data is extracted and stored in XML data format.

The Figure 5 shows the Function Call Trace data for Port Scan malicious JavaScript, converted to XML format.

```
1  <root>
2  <document.write>
3  <P1><div id="comments_thread">Comments.</div></P1>
4  <Loc>Sampe1:14398</Loc>
5  </document.write>
6  <setInterval>
7  <P1>100</P1>
8  <P2>function startRequest() {
9  createXMLHttpRequest();
10 xmlHttp.onreadystatechange = handleStateChange;
11 xmlHttp.open("GET", settingUrl, false);
12 xmlHttp.send();}</P2>
13 <Loc>Sample1:15232</Loc>
14 </setInterval>
15 <XMLHttpRequest.open>
16 <P1>GET</P1>
17 <P2>http://192.168.159.133</P2>
18 <P3>false</P3>
19 <Loc>Sample1:15622</Loc>
20 </XMLHttpRequest.open>
21 <XMLHttpRequest.send>
22 <Loc>Sample1:15665</Loc>
23 </XMLHttpRequest.send>
24 </root>
```

Figure5. Trace Data of a Port Scan Malicious Script

In this article, SimHash Algorithm[4] is proposed for comparison of JavaScript Function Call Trace similarities. SimHash utilizes Local Sensitive Hashing (LSH) for similarity comparison, and LSH maximizes conflicts between similar items rather than avoiding them. That is, the algorithm generates similar results for similar items. Using this function, regardless of the input value size, generate FingerPrint in an array in bit form just like the outcome of a normal hash function, and then use the hamming distance to measure the similarity

## 4. TECHNIQUE OF GENERATING A SIGNATURE DEDICATED FOR DETECTION

In this article, the malicious script, malicious type, obfuscation status, meta data and other data received from the analytics engine are used for automatic generation of signature for malicious JavaScript. It is proposed that a detection signature can be automatically generated by clustering with a malicious script from the registered malicious script pool, generating the combined signature, and refining the signature. Figure 6 illustrates the process of signature generation.

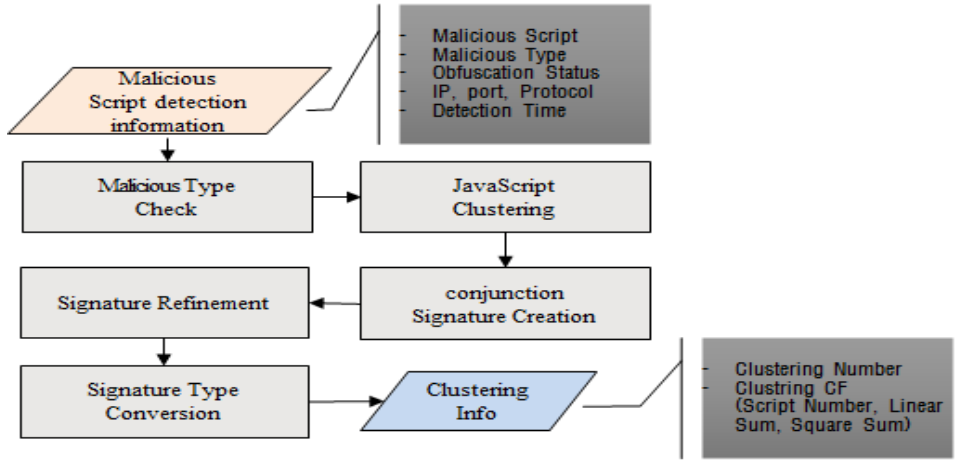

Figure6. Automatic Generation of the Detection Signature for a Malicious Script

## 4.1 Malicious Script Clustering

In this article, it is proposed to use the script clustering technique for automatic generation of the detection signature for a malicious script. The goal of clustering is to streamline the signature itself and improve the false positive rate by grouping malicious scripts showing similar behaviors, and thus preventing extraction of unnecessary tokens. For each token of malicious JavaScript, calculate the Term Frequency-Inverse Document Frequency value and vectorize it. The TF-IDF[5] weight is a statistical figure that is used to evaluate the importance of a certain term in a document, and it can be calculated as the product of Term Frequency and Inverse Document frequency.

The Term Frequency simply indicates how often a term appears in the document, and the Inverse Document Frequency provides general importance of the term.

$$tf_{i,j} = \frac{n_{i,j}}{\sum_k n_{k,j}} \qquad (1)$$

- $n_{i,j}$ indicates the number of times that Term $t_i$ appears in document $d_j$.

$$idf_i = \log \frac{|D|}{|\{d_i : t_i \in d_j\}|} \qquad (2)$$

- $|D|$ indicates Total Document Numbers
- $|d_i : t_i \in d_j|$ indicates number of documents in which term $t_i$ appears

$$tfidf_{ij} = tf_{ij} * idf_i \qquad (4)$$

- TF-IDF weight is calculated by multiplying the TF and IDF.

Using the vector created with TF-IDF, perform hierarchical clustering in the Complete-linkage Cluster method. By improving BIRCH Algorithm for hierarchical clustering, quantify the vector distance and meta data (time similarity, IP, port and protocol), and then take their sum as the similarity score to determine whether malicious script clustering can be done.

Figure 7 shows the clustering process through modified distance calculation. For clustering purpose, the score of significant meta data similarity is applied to distance measurements between basic vectors in order to form a clustering tree.

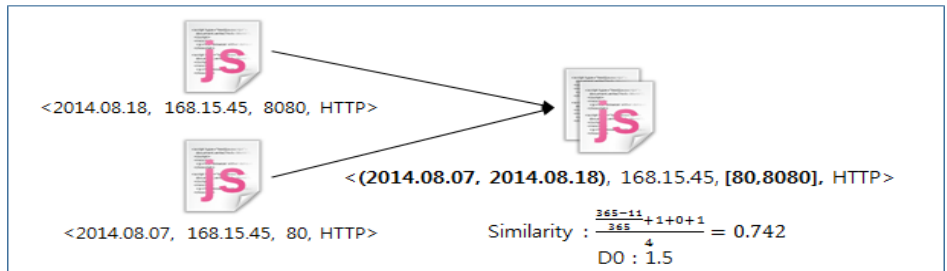

Figure7. Modified Distance Calculation Formula and Meta Similarity Application

### 4.2 Generating a Conjunction Signature

Extract a common token from a malicious script file within the allowed distance in a cluster to generate a conjunction signature. Convert a token in the same form, such as IP, to a regular expression before processing. The Table 3 shows the combined signature generated with Port Scan JavaScript.

Table 3. Examples of Conjunction Signature for Port Scan Detection

output, targetIP, endtime, starttime, ate, appendChild, break, wordWrap, createElement, onRequest, ActiveXObject, majorPort, (?:(?:25[0-5]|2[0-4][0-9]|[01]?[0-9][0-9]?) \ \.) {3 }(?:25[0-5]|2[0-4][0-9]|[01]?[0-9][0-9]?)XMLHttpRequest, style, open, innerHTML, Array, XMLHTTP, true, onreadystatechange, restime, Microsoft, Close, send, scanRes, getElementById

## 5. CONCLUSION

In this paper, we propose three performance indicators for the real time detection and blocking malicious scripts technique of performance measurement that are Real-time traffic throughput, Web access latency, the malicious script detection rate. For performance goal validation test bed was constructed as shown in Figure 8.

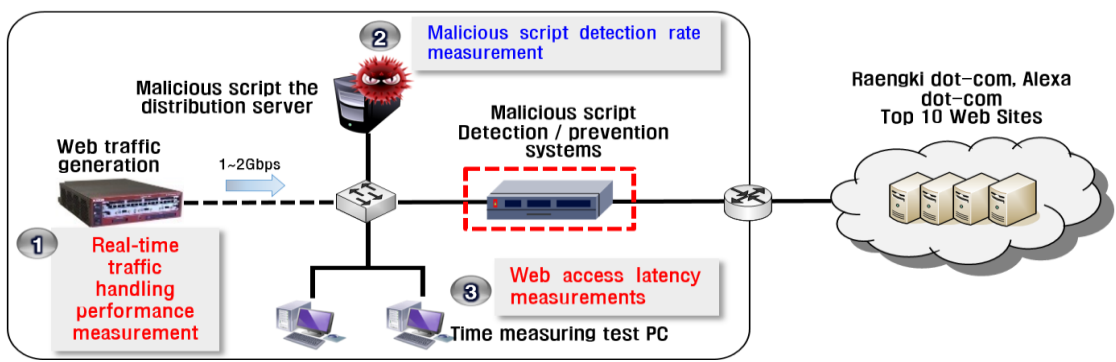

Figure8. Performance Verification Test Environment

The actual run a web browser 20 of the virtual environment at the same time was measured bandwidth and web access latency of about 1Gbps of bandwidth and latency of 2.95 seconds was confirmed that generation.
Figure 9 is a graph showing the traffic handling capacity.

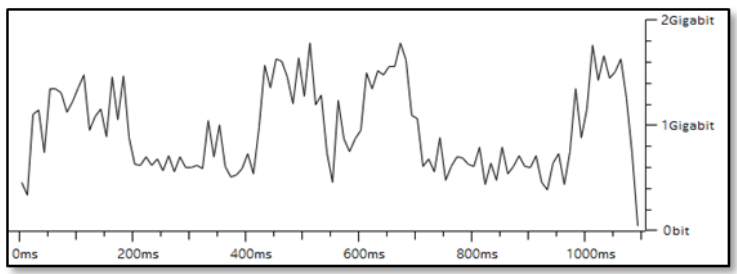

Figure8. Traffic handling performance monitoring

In order to measure the detection rates of malicious script XHR Dos, IP / Port Scan, Web Storage, File API, a total of seven samples of malware types and a total of more than 1400 tests were steady and verify the accuracy of approximately 95.3%. Table 4 shows the results of detecting malicious script.

Table 4. Malicious script detection results

| Malicious Script Type | Code obfuscation attack detection rates | Modified the attack code positives |
|---|---|---|
| XHR Dos | 91%(55/60) | 96%(58/60) |
| IP Scan | 86%(52/60) | 93%(56/60) |
| … | | |
| Geolocation | 88%(53/60) | 100%(60/60) |
| Average | 89% | 97% |

In this article, a proposal has been made for techniques that are used to detect malicious JavaScript and to automatically generate detection signatures. While it shows good results if the signatures generated using the proposed techniques are employed to detect malicious scripts and measure the latency time, it requires additional experiments on a larger pool of samples and higher volume of traffics. Furthermore, to deal with security vulnerabilities of new APIs in HTML5, it is planned to expand the scope of the proposed dynamic analysis and conduct related studies on detection of malicious behaviors by monitoring behaviors caused by JavaScript running,

## 6. ACKNOWLEDGMENT

This work was supported by the ICT R&D Program of MSIP/IITP. [14-912-06-002, The Development of Script-based Cyber Attack Protection Technology]